\documentclass[11pt,twoside]{article}

\usepackage{asp2006}
\usepackage{epsf}
\usepackage{psfig}
\usepackage{lscape}

\begin{document}
\title{Progress Toward a VLBA Movie of the Jet Collimation Region in M87}   

\author{R. Craig Walker}
\affil{NRAO, P.O. Box O, Socorro, NM 87801 USA}
\author{Chun Ly}
\affil{Department of Astronomy, University of California, Los Angeles, 
CA 90095 USA}
\author{William Junor}
\affil{ISR-2, MS-D436, Los Alamos National Laboratory, Los Alamos, 
NM 87545 USA}
\author{Philip E. Hardee}
\affil{Department of Physics and Astronomy, the University of Alabama, 
Tuscaloosa, AL 35487 USA}

\begin{abstract} 

With its high black hole mass, proximity, and bright jet, M87 provides
the best prospect for a direct imaging study of the acceleration and
collimation region of a jet.  Previous VLBI observations have shown an
edge brightened structure with a wide opening angle at small scales
\citep*{jun99}.  An effort to measure component speeds in this region
using existing VLBA data at 43 GHz gave tentative results of $0.25c$
to $0.4c$ but also indicated that faster sampling is needed
\citep*{ly07}.  Here we provide a progress report on a project to make
a properly sampled movie of motions in the inner jet using the VLBA at
43 GHz.  A pilot project during 2006 measured speeds of about
$0.6c$ and was used to set a frame interval of 3 weeks for the movie.
The movie observations began in January 2007.  Results from the pilot
and from the first frame of the movie are presented.  The goal of the
project is to provide observations of the structure and dynamics of
the jet on scales from under 100 to a few hundred Schwarzschild radii
that can be compared with expectations from theoretical studies and
numerical modeling.

\end{abstract}

\section{Introduction}

Simulations of the regions near black holes where jets are launched
suggest that the jet on small scales may include an inner
relativistic jet, that might be in the form of Poynting flux,
surrounded by slower material that might start as a wind from the
accretion disk \citep{Ha06,McK06,Miz06}.  In these scenarios, the jet
structure and launch mechanism are related to magnetic fields that are
wound up by being dragged around the black hole by the disk material.
Near the disk, these fields have a significant component away from the
black hole which will likely cause the jet to have a wider opening
angle near its base than farther out.  The acceleration of the jet
material may be happening throughout this region.  Understanding this
region would be aided greatly if images and movies could be made with
adequate resolution.

Resolving the inner portions of a jet is difficult because of the
small angular sizes involved.  The largest angular size black holes
are the one in the Galactic Center, which does not show a jet, and
those in a few nearby galaxies which have very massive holes.  Even in
these cases, the black hole is only a few microarcseconds in angular
size as seen from the Earth.  This is well beyond the resolution of
any current imaging instruments.  But the interesting effects in the
collimation and acceleration region might be seen over scales of tens to
hundreds of Schwarzschild radii, or tenths of a milliarcsecond (mas),
in the best cases.  This is a regime that can just be reached with
Very Long Baseline Interferometry (VLBI) at high frequencies.

In order to make useful images and movies of the inner jet, one must
observe a source that has a sufficiently strong jet at high radio
frequencies to allow a VLBI instrument to see significant details of
the structure.  One must also trade off the resolution of the very
highest frequencies with the much better imaging capabilities
available at lower frequencies.  Elsewhere in this volume,
\citet{kr07} describe efforts to image jets at 86 GHz and higher
frequencies.  For our project, we have chosen to observe at 43 GHz
where the imaging capabilities of the Very Long Baseline Array 
\citep[VLBA;][]{ napier94}
alone are good.  It would not be practical to observe on a global
array as often as required for a rapidly-evolving, nearby source and
the atmosphere plus the availability of only 8 VLBA antennas at 86 GHz
make the imaging capability at that frequency problematic.

The choice of source to use to study the inner jet is rather clear.
M87, with a $\sim 3\times 10^9 M_{\sun}$ black hole \citep{harms94,
macchetto97, marconi97} and a distance of only 16 Mpc
\citep{whitmore95,tonry01}, has one of the largest angular size black
holes available.  For the parameters above, the Schwarzschild radius
of about 60 au subtends 3.7 microarcseconds.  Of the sources with
large angular size black holes, M87 is the only one with a jet that is
sufficiently strong at high radio frequencies to allow detailed studies 
of the dynamics.  For the declination of M87 ($12\deg$), the
typical resolution of the VLBA at 43 GHz of $0.4 \times 0.2$ mas corresponds
to 110 by 54 Schwarzschild radii or 0.031 by 0.015 pc.  The observed
structure of the M87 jet is very favorable with significant cross-jet
resolution and enough strength to see the jet for more than 30 times
the resolution along the jet.  The basic structure of M87 at 43 GHz
was first published in \citet{jun99}.  That structure includes a very
wide opening angle in the first few tenths of a mas and a pronounced
two sided jet structure.  The wide opening angle suggests that the
collimation region is beginning to be resolved.

The radio jet in M87 has been observed with VLBI many times
\citep[see] [and references therein]{dodson06,kov07,ly07}.  Typically
motions significantly less than the speed of light are seen, including
recent results at 15 GHz reported in this volume where features are
almost stationary \citep{ko07,kov07}.  But at higher resolutions,
the situation is unclear because any fast motions would be seriously
undersampled by the previous observations.  We report here on the status of our
efforts to delineate properly the motions in M87 on scales from around
60 to several hundred Schwarzschild radii using the VLBA.

\section{The Observations} 

\begin{figure}[!t] 
\plotone{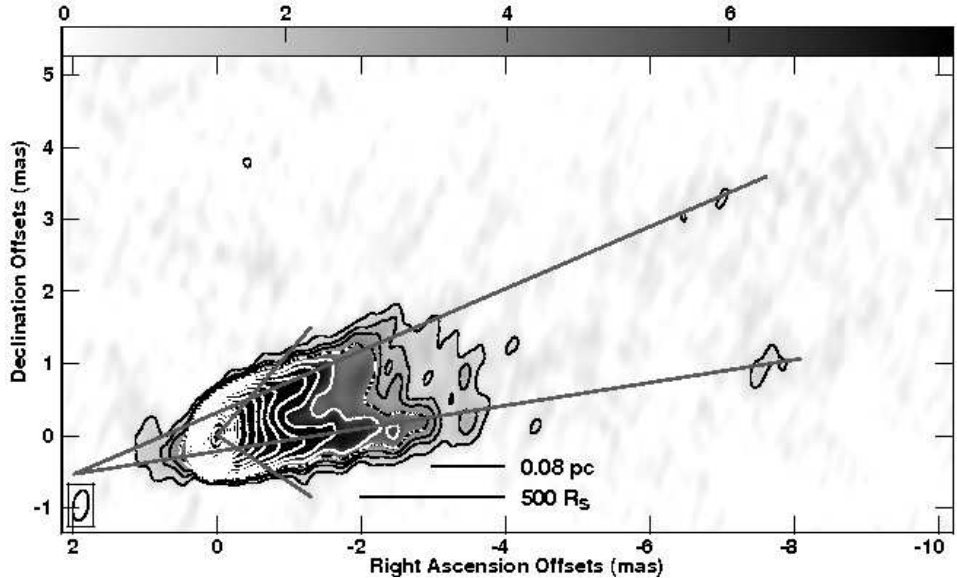} 
\caption{An image made by averaging all five 43\,GHz observations of
\citet{ly07}.  This is a black-and-white version of Figure 3 of that
paper.  The longer lines follow the edge brightened jet emission from
about 1\, to 10\,mas west of the core, with an opening angle of
$15\deg$, while the shorter lines represent the wide opening angle
proposed for sub-mas scales by \citet{jun99} based on slightly higher
resolution Global VLBI data.  Note that the short lines extend beyond
their region of applicability simply to make them visible.  The gray
scale is from 0 to 8 mJy beam$^{-1}$ with contour levels of $-$1, 1,
2, 2.8, 4, 5.7, 8, and multiples of $\sqrt{2}$ thereafter until 512
mJy beam$^{-1}$. The resolution is $0.43 \times 0.21$ mas. Note that
the apparent smoothness of the structure is an artifact of averaging
images from multiple epochs and should not be taken as an indication
that the structure of the jet is featureless. }
\label{yearly}
\end{figure}

Three sets of observations are described here.  The first consists of 5
images at approximately 1 year intervals that were gathered from the
VLBA archive from projects that observed M87 for other purposes
\citep{ly07}.  The structure of the M87 jet was generically similar in
all 5 images, but association of individual features between epochs
was not clear.  The two closest images, separated by about 8 months,
showed some features whose appearance and location suggested that they
could be the same structures seen at both times.  If so, they were 
moving at apparent speeds of $0.25c$ to
$0.40c$.  It would not have been possible to measure such speeds with
any but the closest pair of observations because the motions would
have been too large to allow association of features from one image to
the next.  Of course, these speeds should be treated as lower limits
because a combination of faster speeds and misassociation of
components could give similar rates.  An image, formed from the
average of all 5 observations, is shown in Figure \ref{yearly}.  It
shows the basic structure of the source with more sensitivity than is
available from any one epoch, but any detailed structure seen in
individual epochs is smoothed out.

Another result of \citet{ly07} is the detection of weak emission
opposite the brightest feature from the main jet.  The brightest
feature is presumed to be at, or very close, to the ``core'' --- the
location of the black hole.  In this case, the emission on the other
side would be from a counterjet.  An alternative explanation is that
such emission is actually from the inner jet and that the black hole
is located a couple of mas to the east of the brightest feature.  But
results by \citet{ko07} show that the emission extends beyond the
point where the extrapolation of the edges of the main jet, shown by
the longer lines in Figure \ref{yearly}, converge and has a flaring
structure.  Also there is circular polarization associated with the
brightest feature.  These characteristics suggest that the brightest
feature really is the core and the features to the east really are
associated with a counterjet.  Hence, from here on we will call the
brightest feature the core and the eastern features the counterjet,
even though these words imply a physical structure that is not
completely established.  \citet{ly07} find apparent motions of the
counterjet of about $0.17c$ away from the core.  The motions and
relative brightness of the jet and counterjet could be explained if
the actual jet speed is $0.3-0.5c$ and the angle to the line of sight
is $30\deg$ to $45\deg$.  That would be consistent with some earlier
results, but is hard to reconcile with the observation of superluminal
motions seen in knot HST-1 \citep{ha07} and other features further out
\citep{bi99}, which, by the usual end-on jet model, require small
angles to the line-of-sight.

M87 is sufficiently close that fast motions lead to large angular
rates.  In fact, for motions of $6c$, as has been reported further
down the jet \citep{bi99}, the angular rate would be about 24 mas
yr$^{-1}$!  The VLBA observations at 43 GHz have a resolution along
the jet of about 0.2 mas, so such motions would be 120 beam widths in
a year or about a beam width every 3 days.  Clearly observations on
the order of once per year would be woefully inadequate to measure
such speeds with this resolution.  So we began a project to attempt to
pin down the rates near the core and to study details of the motions.
The observational goal of the project is to produce a movie of the
activity near the core.  But the first task had to be to determine the
appropriate frame rate for the movie.

\begin{figure}[!t]
\plotone{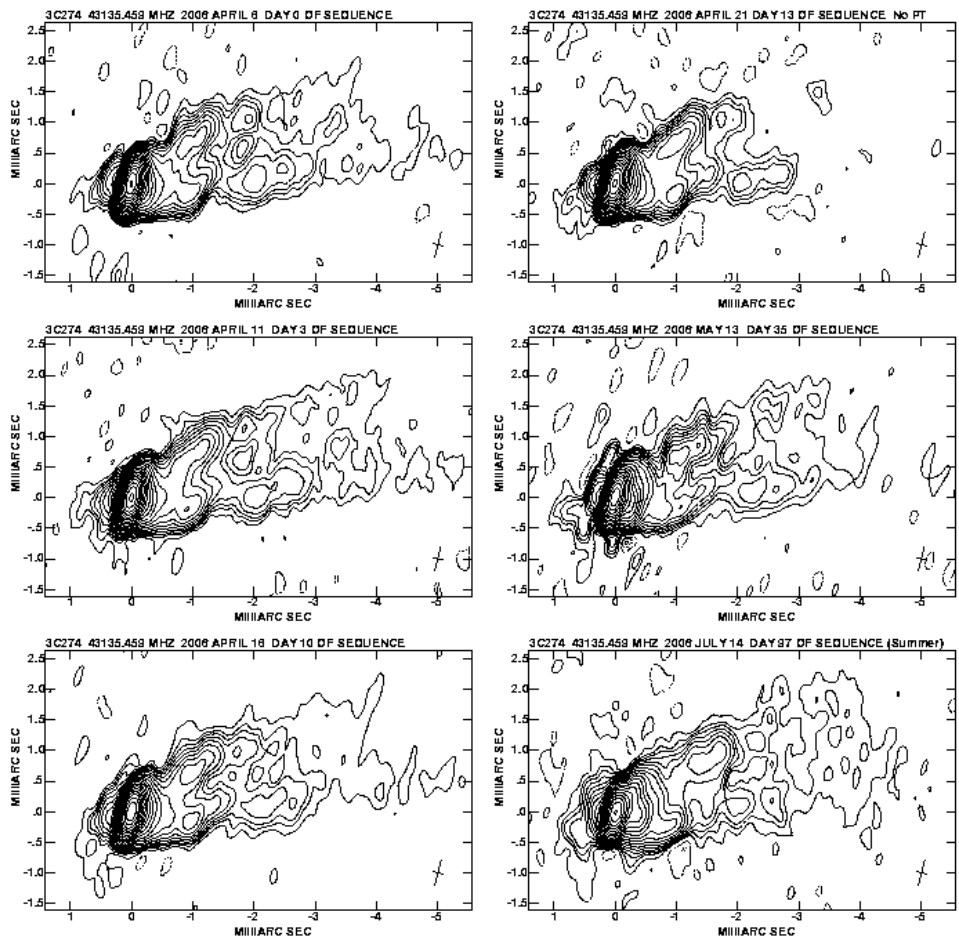}
\caption{Preliminary pilot project images of M87 at 43 GHz made using
the VLBA.  The resolution (FWHM) is $0.43 \times 0.18 $ mas in PA $-14\deg$
as indicated by the crosses in the lower right of each panel.  The
contour levels are $-1$, 1, 1.9, and 2.7 mJy beam$^{-1}$ and increasing
logarithmically thereafter with 7 contours per decade.  As noted in
the individual headers, the observations occurred on days 0, 3, 10,
13, 35, and 97 after the first.  The fourth image has reduced
sensitivity to larger structures because Pie Town, an inner station,
was missing.  The final image has somewhat reduced quality because it
was observed in summer conditions when rapid phase fluctuations are a
problem.}
\label{pilot}
\end{figure}

During the first half of 2006, six observations were made of M87 with
the VLBA at 43 GHz with intervals ranging from 3 days to about 3
months.  The images are shown in Figure \ref{pilot}.  The images made 3 days
apart look very similar.  Apparently there are no superluminal speeds
in this region.  Closer study of the images led to the conclusion that
the speeds of the major features are on the order of 2.2 mas yr$^{-1}$
at 1.5 mas from the core.  That corresponds to about $0.6c$, not much
faster than what was observed from the archive data.  That speed
corresponds to about one beam width per month.  Based on these results,
a frame interval of 3 weeks was chosen for the movie as a compromise between
attempting to avoid undersampling the motions and keeping the data 
volume manageable.

The pilot project images continue to show the double-sided jet
structure as seen previously.  They also show the presumed counterjet.
There do appear to be changes in the structure of the counterjet, but
the nature of those changes is still unclear and an understanding
awaits the movie results.  So far, one of the pilot observations has
been reduced to show the polarization structure, but the only
believable polarized emission observed is along the downstream side of
the core.  An understanding of the polarization structure awaits much
more work with the pilot and movie data, and possible stacking
of the various epochs.

\begin{figure}[!t]
\plotone{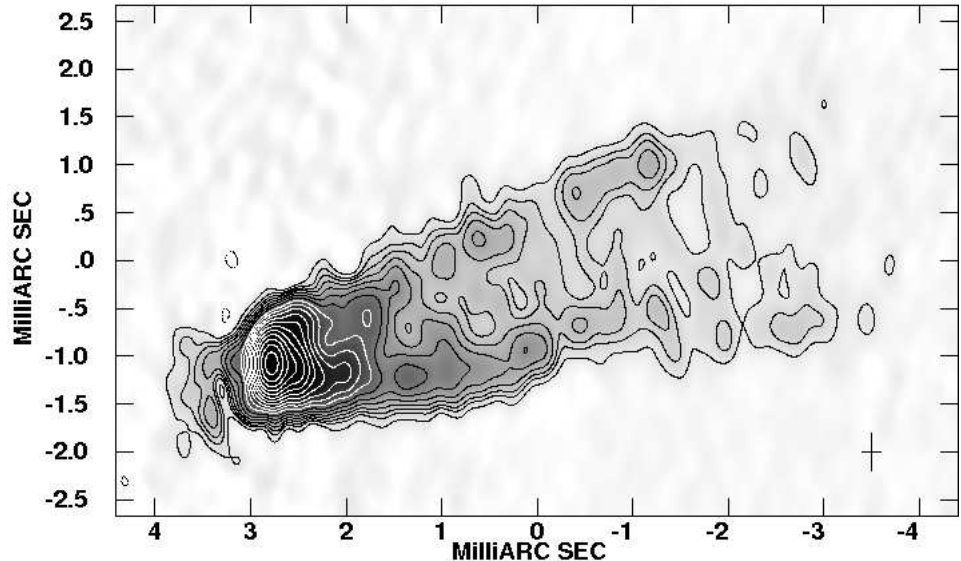}
\caption{The first image of the VLBA movie of M87 at 43 GHz.  The 
resolution is $0.40 \times 0.20$ mas in PA $0\deg$ as indicated by the cross
in the lower right.  The contour levels are $-1$,
1, 2, 2.8, and 4 mJy beam$^{-1}$ and increasing from there by factors 
of $\sqrt{2}$.  The observations were made on 27 January 2007.}
\label{movie01}
\end{figure}

The movie observations began in January of 2007 and are scheduled to
be completed in January 2008.  The bandwidth (a bit rate of 256 Mbps)
is double that used in the pilot and other previous observations so
the sensitivity is improved, allowing the jet to be seen farther from
the core.  The image from the first movie frame is shown in Figure
\ref{movie01}.  The counterjet is still present, and shows an
interesting structure that is well resolved north-south.  And the main
jet structure still has the previously-observed edge-brightened
character.  In the pilot, the northern edge was somewhat brighter than
the southern edge, especially in the $1-2$ mas range from the core,
although in the final pilot image, a bright feature was appearing on
the southern side.  In the first movie images, the southern side has
become dominant in that range.  It is clear from this and the archive
project that the jet edges are not balanced in brightness, but that
the dominant edge switches as new features appear.

As of the time of the meeting, only two frames of the movie had been
reduced.  They showed motions, but again the nature of what was going
on was somewhat confusing and an understanding awaits addition of
several more frames.  Consistency between observations, or lack
thereof, is one of the most powerful indicators of the reliability of
features and motions.

\section{Conclusion}

An effort is well under way to delineate the motions and dynamics of
radio structures on scales of 60 to hundreds of Schwarzschild radii
in M87 using the VLBA at 43 GHz.  Archival data at roughly one year
intervals showed that structural changes do occur, but the sampling
was too coarse to understand the nature of those changes.  A pilot
project was conducted in 2006 to determine roughly the rates of any
motions on the scales of interest to allow the frame rate for a proper
movie to be set.  A frame interval of 3 weeks was selected and a one-year
sequence of 18 frames is in progress.  A mini-movie with the first two
frames was shown at the meeting.  A full delineation of any motions
near the core of M87, including an understanding of why the speeds
measured by the pilot project are apparently inconsistent with the low
rates seen by \citet{kov07}, awaits more complete results from the
movie.



\acknowledgements 

This research has made use of the NASA Astrophysics Data System.
W. Junor was partially supported through NSF AST 98-03057.  C. Ly
acknowledges support through the NRAO Graduate Research Program.
P. Hardee is supported by NASA/MSFC cooperative agreement NCC8-256 and
NSF award AST-0506666 to the University of Alabama.  The National
Radio Astronomy Observatory is a facility of the National Science
Foundation, operated under cooperative agreement by Associated
Universities, Inc.

\end{document}